\documentclass{article}

\usepackage[left = 2cm, right = 2cm, top = 2cm, bottom = 2cm]{geometry}

\usepackage[comma, sort&compress]{natbib}

\usepackage[utf8]{inputenc}
\usepackage{authblk} 
\usepackage{setspace} 
\usepackage[english]{babel} 
\usepackage{mathpazo} 
\usepackage{amsmath, amsthm, mathtools, amsfonts} 
\usepackage{lineno} 
\usepackage[table]{xcolor} 
\usepackage{longtable, lscape} 
\usepackage[colorlinks=true,allcolors={blue!80!black}]{hyperref} 
\usepackage[capitalise, noabbrev]{cleveref} 
\usepackage{multirow} 
\usepackage[title,titletoc]{appendix} 

\usepackage{tikz}
\usetikzlibrary{matrix, calc, shapes.geometric, arrows}
\tikzstyle{outcome} = [rectangle, rounded corners, minimum width = 3cm, minimum height = 1cm, text centered, text width = 3cm, draw = black]
\tikzstyle{reason} = [rectangle, rounded corners, minimum width = 10cm, minimum height = 0.75cm, text width = 10cm, draw = black]
\tikzstyle{arrow} = [thick, ->, >= stealth]

\usepackage{caption} 
\usepackage{subcaption} 
\usepackage{graphicx} 
\usepackage{rotating} 

\makeatletter
\def\hlinewd#1{%
  \noalign{\ifnum0=`}\fi\hrule \@height #1 \futurelet
   \reserved@a\@xhline}
\makeatother

\newcommand{\inla}{\texttt{r-inla}}
\newcommand{\summer}{\texttt{SUMMER}}

\DeclarePairedDelimiter\abs{\lvert}{\rvert} 
\newcommand{\pa}[1]{\left(#1\right)} 
\newcommand{\bra}[1]{\left[#1\right]} 
\newcommand{\rOpenInt}[1]{\left[#1\right)} 
\newcommand{\bs}[1]{\boldsymbol{#1}} 
\newcommand{\logit}{\text{logit}} 
\newcommand{\expit}{\text{expit}} 
\newcommand{\II}{\mathbb{I}}

\graphicspath{ {./figures/} }

\defcitealias{KDHS2014}{KDHS, 2015}

\title{Estimating Subnational Under-Five Mortality Rates Using a Spatio-Temporal Age-Period-Cohort Model}

\author[1*]{C. Gascoigne}
\author[2]{T. Smith}
\author[3]{J. Paige}
\author[4,5]{J. Wakefield}

\affil[1]{MRC Centre for Environment and Health, Department of Epidemiology and Biostatistics, School of Medicine, Imperial College London, London, UK}
\affil[2]{Department of Mathematical Sciences, University of Bath, Bath, UK}
\affil[3]{Department of Mathematical Sciences, Norwegian University of Science and Technology, Trondheim, Norway}
\affil[4]{Department of Statistics, University of Washington, Seattle, WA, USA}
\affil[5]{Department of Biostatistics, University of Washington, Seattle, WA, USA}
\affil[*]{Corresponding author. Email: \href{email:email-id.com}{c.gascoigne@imperial.ac.uk}}

\date{}

\begin{document}

\maketitle

\abstract{Producing subnational estimates of the under-five mortality rate (U5MR) is a vital goal for the United Nations to reduce inequalities in mortality and well-being across the globe. There is a great disparity in U5MR between high-income and Low-and-Middle Income Countries (LMICs). Current methods for modelling U5MR in LMICs use smoothing methods to reduce uncertainty in estimates caused by data sparsity. This paper includes cohort alongside age and period in a novel application of an Age-Period-Cohort model for U5MR. In this context, current methods only use age and period (and not cohort) for smoothing. With data from the Kenyan Demographic and Health Surveys (DHS) we use a Bayesian hierarchical model with terms to smooth over temporal and spatial components whilst fully accounting for the complex stratified, multi-staged cluster design of the DHS. Our results show that the use of cohort may be useful in the context of subnational estimates of U5MR. We validate our results at the subnational level by comparing our results against direct estimates.}

\section{Introduction}

The United Nations (UN) estimate under-five mortality rates (U5MR) at a subnational level in part to assess their Sustainable Development Goals (SDGs) target 3.2: ``By 2030, end preventable deaths of new-borns and children under five years of age, with all countries aiming to reduce neonatal mortality to at least as low as 12 deaths per 1,000 live births and under-five mortality to as least as low as 25 per 1,000 live births'' \citep{sdgsWeb}. Interventions normally take place at a local authority level; therefore, estimation and prediction of U5MR at a subnational level are important goals of the UN to ensure the maximum impact and cost effectiveness of any future intervention. For example, the optimal intervention may differ region-to-region, so nationwide interventions are less efficient in terms of time, impact, and cost. 

The U5MR from developed countries, such as those from Europe and North America, are not of great concern to the UN since these are typically much lower than those from, say, sub-Saharan Africa, where children are 14 times more at risk of dying in the first five years of life \citep{igme2021}. Due to this, the UN focus their attention on finding U5MR estimates for low-and-middle income countries (LMICs) to reach  SDG 3.2 target. It is common for LMICs to not have vital registration systems (registration of all births and deaths) with the required coverage. The most reliable source of data are household surveys such as the Demographic and Health Surveys \citep[DHS;][]{dhs}. The DHS has conducted more than 400 surveys in over 90 countries and has collected data that can be used for national (Admin-0), county (Admin-1) and local authority (Admin-2) level inference. However, leveraging spatial correlation is required for finer geographical resolutions.

Small area estimation (SAE) techniques are used to estimate U5MR for geographical regions where there are little to no samples available. Here we briefly review the main techniques, but for a comprehensive review of SAE in this context, see \cite{wakefield2020small}. An important quantity to calculate in the context of spatial modelling of U5MR using SAE techniques is the survey design weighted (direct) estimates \citep{horvitz1952generalization}. Each individual has an associated weight which is proportional to the inverse inclusion probability to account for the sampling design, for example, the DHS uses a stratified two-staged cluster sampling design. Direct estimates are considered the gold standard for large samples since they are design consistent when the weights are reliable. This means that as the number of people who are included in the samples gets closer to the total population, the direct estimate will converge to the true population value. However, when data is sparse, such as on a subnational level, the direct estimates suffer from a large amount of sampling variability, and the variance for these methods becomes unacceptably large.

Considering the problem of data sparsity, \cite{fay1979estimates} introduced a model that uses a (possibly) transformed version of the direct estimates to gain precision by smoothing the direct estimates over time using a random effects model. The Fay-Herriot (FH) approach captures the concept that, in general, we expect observations that are closer in time are more likely to be like one another than distant observations. By incorporating this idea using a random effects model, the direct estimates can borrow strength from neighbouring areas to calculate an estimate with a smaller variance. We call these estimates the FH estimates. 

The direct and FH estimates capture the complex survey design using design weights. Methods such as cluster level models do not use design weights but attempt to capture the complex design through other methods, i.e., including survey design specific covariates in the model and by including overdispersion parameters to acknowledge cluster. An example of cluster-level analyses is that of \citet{wakefield2019estimating} and \citet{li2020space}, which combines a discrete time survival analysis (DTSA) approach with space-time models to estimate U5MR which have been smoothed over trends in space (location) and time.

In cluster level models, the age at death, the year of death and the spatial location of where the death occurred have been taken into consideration, but the cohort (year of birth) has not. This is the same for the direct and FH estimates, and, to the best of our knowledge, cohort has not been considered in the context of U5MR. A cohort effect is a measure of long-term exposures that those born at a similar time encounter together as they move through life. Often smoothing across trends in cohort is performed to aid in the production of stable estimates and prediction \citep{knorr2001projections}. Specifically including cohort alongside age has been shown to produce better out-of sample predictions when compared to a model that uses age and period for mesothelioma cancer predictions \citep{martinez2016simple}. Furthermore, in the context of pre-mature (early adulthood) all-cause mortality, the inclusion of long-term cohort trends alongside period was shown to produce predictions less influenced by sudden changes that may not be representative of the overall trend \citep{best2018premature}. For sparse data situations, such as those for subnational estimates of U5MR from the DHS, smoothing is necessary to produce reliable estimates and measures of uncertainty for policymakers to make vital decisions about interventions to attain goals such as those included in SDG 3.2. Therefore, the inclusion of cohort aligns with the goals of subnational modelling of U5MR and would be a natural inclusion in this setting. 

In this article, we modelled the Kenyan U5MR at a subnational level with the inclusion of effects for (birth) cohort alongside those for age and period (year of death), we opted to use an age-period-cohort (APC) model. We chose Kenya since it is a well modelled country from the DHS and has been used previously in the context of U5MR \citep{ettarh2012determinants, macharia2019sub, wakefield2019estimating, macharia2021subnational}. In U5MR models, age is typically the most important temporal trend due to how rapidly mortality changes in the earlier months. In developing countries, the first month of life can account for almost 50\% of all under-five deaths \citep{igme2021}, and in the 2014 Kenyan DHS \citepalias[KDHS;][]{KDHS2014}, deaths in the first month of life account for 42\% of the total number of deaths. The period (year) effect is a measure of short-term exposures, such as a new treatment and the (birth) cohort effect is a measure of longer-term exposures. APC models have been used in several different health concerns such as suicide rates \citep{riebler2012gender}, stomach cancer incidence \citep{papoila2014stomach}, all-cause mortality \citep{smith2018stratified} and opioid overdose mortality \citep{chernyavskiy2020spatially}. APC models have a well-known identification problem due to the linear dependence between the three temporal effects ($cohort = period - age$). Many authors have addressed this, and there are several appropriate approaches to produce estimates that are invariant to the choice of parameterisation \citep{holford1983estimation, clayton1987b}.  

An APC model to produce subnational estimates of U5MR is interesting to consider because current methods such as direct estimates \citep{horvitz1952generalization}, FH estimates \citep{fay1979estimates}, and other cluster level models \citep{wakefield2019estimating, li2020space} do not consider a cohort effect, which has been found helpful in numerous health scenarios where smoothing is required for estimation and prediction. In Section 2, we introduce the data used. In Section 3, we detail the APC model with specific extensions to account for stratification and clustering, as well as the method to calculate U5MR. Section 4 contains the results from the model validation of the APC model alongside the results from models containing only age and period and age and cohort terms. Section 5 is the results of the APC model at the subnational (Admin-1) level and when aggregated to the national (Admin-0) level. Finally, Section 5 concludes the paper.

\section{Data}

A typical DHS survey is a stratified two-stage cluster sampling scheme with stratification by county crossed with an urban/rural indicator drawn from an existing sample frame (often the most recent census frame), which is a complete list of all sampling units that (hypothetically) cover the target population. After deciding on how many primary sampling units (PSUs) to sample within each stratum, the first stage of sampling is to select the PSUs with probability proportional to size sampling. These PSUs are called enumeration areas (EAs) and form the survey clusters. In the second stage, a fixed number (typically 25-30) of households are selected by equal probability sampling from a list of all households in each EA. 

We used the 2014 KDHS and produced yearly estimates for the U5MR for each of Kenya's 47 (Admin-1 regions) counties. The 2014 KDHS is stratified by the 47 counties each with its own urban/rural indicator. The total number of strata was 92 since both Nairobi and Mombasa are entirely urban. In the first sampling stage, 1,612 clusters were selected out of a possible 96,251 (as defined from the 2009 Kenya Population and Housing Census) of which 617 are urban and 995 are rural; urban areas are over sampled. In the second stage of sampling, 40,300 households are sampled from the selected EAs (25 households per EA). 

\begin{figure}[!h]
    \centering
    \caption{Urban and rural cluster locations on the 2014 Kenyan DHS with the 47 county boundaries.}
    \label{Fig: Kenya_adm1_cluster_location_plot}
    \includegraphics[width=0.7\textwidth]{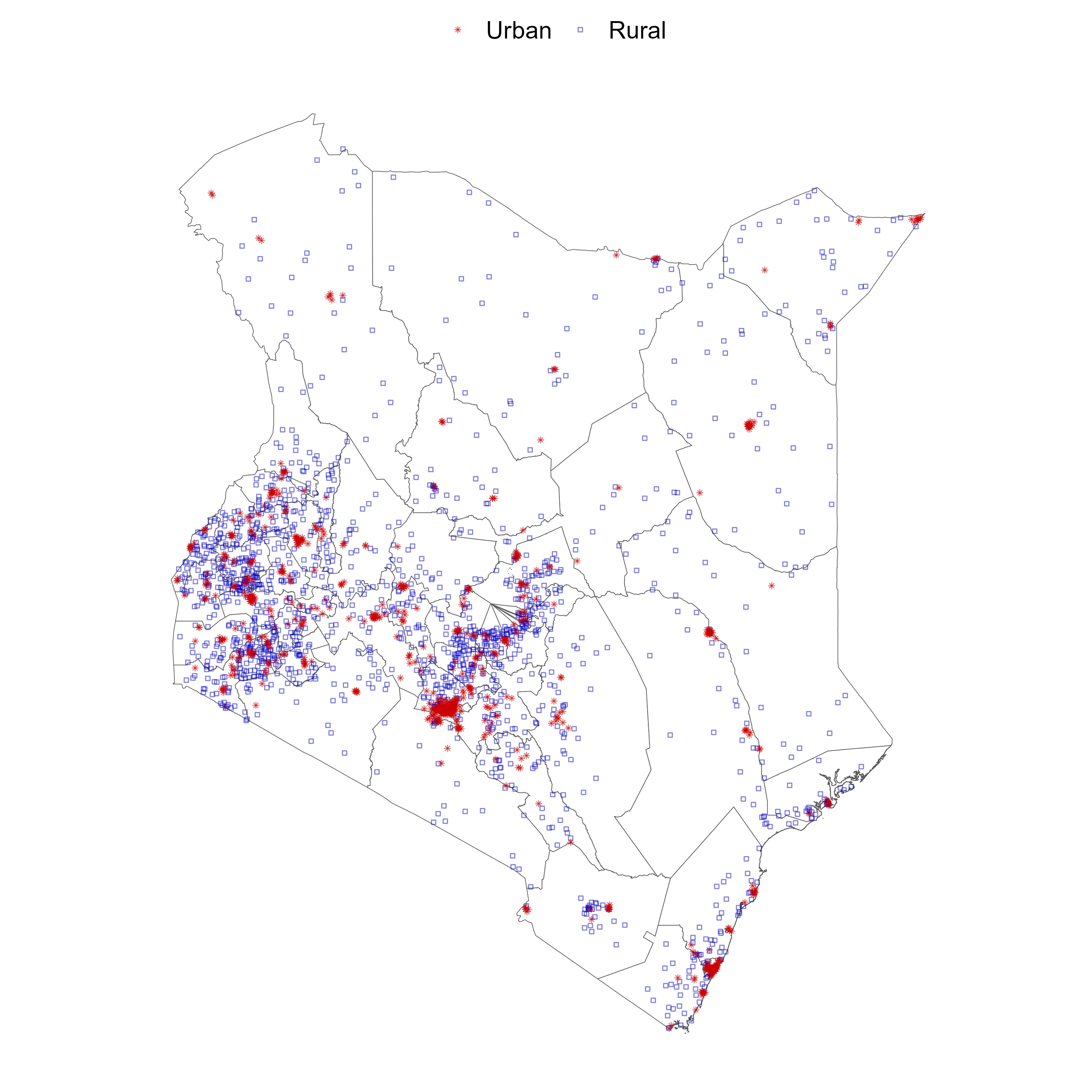}
\end{figure}

Figure \ref{Fig: Kenya_adm1_cluster_location_plot} shows a map of Kenya including county (Admin-1) borders and cluster locations, reflecting the population density and the urban/rural stratification. For confidentiality, the GPS coordinates for cluster centres are jittered with urban and rural coordinates displaced by up to 2 and 5km, respectively. In addition, the locations for a further 1\% of rural locations are displaced by up to 10km.

The data we used to estimate U5MRs were from the DHS birth history questionnaire results. The birth history questionnaire is answered by all women aged 15-49 who spent the previous night in the sampled household and contains information on pregnancy, postnatal care, immunisation, and health of children born in the last five years. Specifically, we concerned ourselves with answers relating to the children that had information on age at death (if occurred), period (year) of death (if occurred) and year of birth (cohort). In addition, the questionnaires contain the survey design information such as the cluster identification, region, and urban/rural stratification. Each row of the survey relates to an individual child. The retrospective nature of the data allows historic trends to be estimated and we use yearly periods between 2006 -- 2014 and monthly ages between 0 -- 60 months.

\section{Methods} \label{Section: Methods}

\subsection{Discrete time survival analysis}

We modelled U5MR using discrete time survival analysis (DTSA) as it allows for flexible modelling of (potentially) complicated temporal relationships; an important feature due to the substantially different hazards into the first five years of life \citep{clark2013young}. To understand how DTSA is used to attain estimates of U5MR, consider the case of a general DTSA. 

Let $A$ be the interval of event (in our case, death) where $A \in \{0, \dots, 59\}$ for the first 59 months of life (i.e., all those months under-five). The hazard $\text{h}$ for month $a$ is defined as the probability of event (death) occurring in the month interval $a$, given it has not occurred before month $a$
\begin{equation*}
    \label{Eq: Hazard Probability}
    \text{h}\pa{a} = p\pa{A = a | A \geq a}.
\end{equation*}
Letting $\text{Survival}\pa{a}$ be the probability of surviving beyond month $a$,
\begin{equation*}
    \label{Eq: Survival Probability}
    \begin{split}
        \text{Survival}\pa{a} & = p\pa{A > a} \\
        & = p\pa{A > a | A \geq a} p\pa{A > a - 1 | A \geq a - 1} \times \dots \times p\pa{A > 1 | A \geq 1} \\
        & = \bra{1 - \text{h}\pa{a} } \bra{1 - \text{h}\pa{a-1} } \times \dots \times \bra{1 - \text{h}\pa{1}}.
    \end{split}
\end{equation*}
where each conditional is the survival probability for that month given they have survived up to the start of the month which is equivalent to the compliment of the hazard probability. As U5MR is the compliment of the survival probability up to and including the $59^\text{th}$ month, it can be written as $1 - \text{Survival}\pa{a = 59}$.

Following \cite{mercer2015small} and \cite{wakefield2019estimating} and for reasons of parsimony, we grouped the months into six discrete hazards (age groups) $\rOpenInt{0,1}, \rOpenInt{1,12}, \rOpenInt{12,24}, \rOpenInt{24,36}, \rOpenInt{36,48}$, and $\rOpenInt{48,60}$ and attributed each month to one of the six hazards using
\begin{alignat*}{2}
    \text{h}_{48} & = \dots = \text{h}_{59} && = \text{h}_{52.5} \\
    \text{h}_{36} & = \dots = \text{h}_{47} && = \text{h}_{41.5} \\
    \text{h}_{24} & = \dots = \text{h}_{35} && = \text{h}_{29.5} \\
    \text{h}_{12} & = \dots = \text{h}_{23} && = \text{h}_{17.5} \\
    \text{h}_{1} & = \dots = \text{h}_{11} && = \text{h}_{6} \\
    \text{h}_{0} & && = \text{h}_{0}.
\end{alignat*}
where $\text{h}_{a} = \text{h}\pa{a}$ was used for notational convenience, $a = 0, \dots, 59$ are the monthly under-five ages, and $\tilde{a} = 0, 6, 17.5, 29.5, 41.5, 52.5$ are the midpoint points of each discrete hazards. With each month attributed to one of the six discrete age groups, the U5MR is
\begin{align}
    \text{U5MR} & = 1 - \text{Survival}\pa{a = 59} \\
    & = 1 - \bra{1 - \text{h}_{59}} \bra{1 - \text{h}_{58}} \times \dots \times \bra{1 - \text{h}_{0}} \\
    & = 1 - \left(\bra{1 - \text{h}_{52.5}}^{12} \bra{1 - \text{h}_{41.5}}^{12} \bra{1 - \text{h}_{29.5}}^{12} \bra{1 - \text{h}_{17.5}}^{12} \bra{1 - \text{h}_{6}}^{11} \bra{1 - \text{h}_{0}} \right) \nonumber \\
    & = 1 - \prod_{i = 1}^{6} \bra{1 - \text{h}_{\tilde{a}\bra{i}}}^{z\bra{i}}
    \label{Eq: Simplified U5MR Hazard Probability}
\end{align}
where the term $\tilde{a}\bra{i}$ reads ``the $i^\text{th}$ element of vector $\tilde{a}$'', and $\tilde{a} = 0, 6, 17.5, 29.5, 41.5, 52.5$ and $z = 1, 11, 12, 12, 12, 12$ are the midpoints, and the number of grouped ages in each of the six discrete hazards, respectively.

With the discrete hazards defined, for each child, we defined a sequence of Bernoulli random variables for the death in given month. This yielded a sequence (potentially of length 60) of binary outcomes 0/1 for survived/died that was attributable to a particular cluster, region, strata, year, and cohort. In this format we had information on when each child survived/died as well as how many months at risk they contributed to. For example, a child who dies after 15 months contributed one, eleven and three months at risk to the $\rOpenInt{0,1}$, $\rOpenInt{1, 12}$ and $\rOpenInt{12, 24}$ age groups, respectively, up to and including the month of death. This process was repeated for all individuals and then we summed over the number of deaths and months at risk for each cluster, region, strata, year and cohort combination.

\subsection{Spatio-temporal APC model}

We used APC and SAE methods to model and analyse the temporal trends and account for stratification and clustering in the survey design, respectively, when estimating the U5MR. In this paper, we combined the APC ideas of \cite{gascoigne2023penalized} with the SAE model of \cite{wakefield2019estimating, li2020space} and defined an APC model with spatial components.

The number of observed deaths and number of months at risk for age group $\tilde{a}$, period $p$, cohort $c$ and cluster $k$ are $y_{\tilde{a},p,c,k}$ and $n_{\tilde{a},p,c,k}$, respectively. As we used the true birth cohorts, there were instances where we had multiple cohorts for a single age and period combination, rather than just one if they were calculated using $cohort = period - age$; therefore, we explicitly denoted $c$. However, we did not explicitly denote region $r$ as this was included implicitly within the cluster index $k$. We use the term region interchangeably with county, but in general, this can be any spatial resolution, i.e., national (Admin-0), county (Admin-1), and local authority (Admin-2). To model the monthly probability of death conditional on the child being alive at the start of a month, we used the overdispersed binomial, cluster-level model
\begin{equation*}
    \label{Eq: BetaBinomial Cluster-Level Model}
    y_{\tilde{a},p,c,k} | \pi_{\tilde{a},p,c,k}, d \sim \text{BetaBinomial}\pa{n_{\tilde{a},p,c,k}, \pi_{\tilde{a},p,c,k}, d}
\end{equation*}
where $\pi_{\tilde{a},p,c,k}$ was the monthly hazard for age group $\tilde{a}$, period $p$, cohort $c$ and cluster $k$. For clarity, $\pi = h$ in Equation \ref{Eq: Simplified U5MR Hazard Probability}. In a beta-binomial model, $d$ is the overdispersion (excess binomial variation) parameter. We included a parameter to account for overdispersion as it is common when modelling health and demographic data. In the context of DHS data, the overdispersion parameter accounts for the clustering aspect of the survey design (i.e., the dependence between births from the same mother, respondents from the same households, and dependence within the same cluster).

To reflect the over and under sampling of urban and rural areas common in the DHS survey data, we included a binary stratification variable for the urban/rural classification. This was necessary as oversampling in urban or rural clusters can lead to bias due to different hazards for each stratum. Alongside the stratification, we included terms for the temporal, spatial and spatio-temporal effects. Using the standard logit transform, the spatio-temporal APC model is
\begin{equation}
    \label{Eq: Reparameterised Spatio-Temporal APC}
    \logit \pa{\pi_{\tilde{a},p,c,k}} = \beta_1 + I\pa{\bs{s}_k \in \text{ urban }}\beta_2 + {t_{1, r}}{\beta_3}  + {t_{2, r}}{\beta_4} + {\nu_{\tilde{a}}} + {\eta_{p}} + {\xi_{c}} + S_{r\bra{s_k}} + \delta_{p,r\bra{s_k}}
\end{equation}
where $I\pa{\bs{s}_K \in \text{ urban }}$ is one if cluster $k$ at location $\bs{s}_k$ is urban so that $\beta_1$ is the intercept for rural clusters and $\beta_1 + \beta_2$ is the intercept for urban clusters. The terms $t_{1,r}$ and $t_{2,r}$ are region-specific temporal linear trends, and ${\nu_{\tilde{a}}}$, ${\eta_{p}}$ and ${\xi_{c}}$ are curvature terms for age, period and cohort, respectively. The spatial and space-period random effects are denoted $S_{r\bra{s_k}}$ and $\delta_{p,r\bra{s_k}}$, respectively. For any term relating to a region $r$, the notation ${r\bra{s_k}}$ reads ``the region $r$ within which cluster $s_k$ resides''.

Including the three temporal effects (age, period, and cohort) together in one model causes the well-known identification issue \citep{clayton1987b, osmond1989age} due to the linear dependency between the three (with two, the third can be found). In addition to the identification issues due to the linear dependency between the temporal linear trends, when data is aggregated in non-equal widths (i.e., the age and period intervals are not of the same length, like the monthly ages and yearly periods used in the context of U5MR), there are additional identification problems that cause a saw-tooth pattern in the estimates of period and cohort \citep{holford2006approaches, riebler2010analysis}. Identification problems relating to the linear dependency were alleviated by reparameterising the temporal terms into a set of identifiable curvatures (that are orthogonal to their respective linear trends) and an arbitrary two (out of three) temporal linear trends \citep{holford1983estimation}. As we were not interpreting the age, period and cohort effects, the choice of what linear trends to retain (${t_{1,r}}$ and ${t_{2,r}}$) was unimportant as the estimates of $\logit\pa{\pi_{\tilde{a},p,c,k}}$ are invariant to this choice. For transparency, we retain the age and period linear trends. The additional identification problems relating to the non-equal widths were alleviated by including a penalty on the second derivative of the curvature terms \citep{gascoigne2023penalized}, which in a Bayesian paradigm is achieved using a Random Walk Two (RW2) prior \citep{rue2005gaussian} for each of the curvature terms. 

In Equation \eqref{Eq: Reparameterised Spatio-Temporal APC}, we included a space-period interaction term to be consistent with the literature \citep{knorr2000bayesian}, but there are other interaction terms that could have been considered. For example, for spatio-temporal interactions there are the age-space and cohort-space interactions and for within temporal interactions there are the age-period, age-cohort, and period-cohort interaction terms. We chose not to include these in this article, but instead save this for future work.

\subsection{Additional models}

Alongside the APC model, we explored the use of age-period (AP) and age-cohort (AC) models for subnational estimation and prediction of U5MR. Both the AP and AC models are within the APC model hierarchy \citep{clayton1987b}, and are commonly compared alongside the results of an APC model. These simpler models do not suffer from the structural link identification problem, meaning we did not need to drop one of the slopes; however, we chose to retain the reparameterisation in terms of curvatures for comparability. The AP and AC models are
\begin{alignat*}{2}
    \text{AP:} \quad & \logit  && \pa{\pi_{\tilde{a},p,c,k}} = \beta_1 + I\pa{\boldsymbol{s}_k \in \text{ urban }}\beta_2 + {\tilde{a}}_{r}{\beta_3} + {p}_{r}{\beta_4} + {\nu_{\tilde{a}}} + {\eta_p} + S_{r\bra{s_k}} + \delta_{p,r\bra{s_k}} \\ 
    \text{AC:} \quad & \logit && \pa{\pi_{\tilde{a},p,c,k}} = \beta_1 + I\pa{\boldsymbol{s}_k \in \text{ urban }}\beta_2 + {\tilde{a}}_{r}{\beta_3} + {c}_{r}{\beta_4} + {\nu_{\tilde{a}}} + {\xi_c} + S_{r\bra{s_k}} + \delta_{p,r\bra{s_k}}. 
\end{alignat*}

The AP model is similar to the cluster level smoothing model of \citet{wakefield2019estimating} and \citet{li2020space}. These models are similar, instead of equal, as they have the same goal, use representations of age and period to produce estimates and predictions of U5MR, but achieve this in different ways. In the AP model, we reparameterised age and period into a linear trend and their orthogonal curvatures where we smoothed over the estimates of age and period curvature. For the cluster model, age is considered as a categorical variable with one level for each of the distinct age groups $\rOpenInt{0,1}, \rOpenInt{1,12}, \rOpenInt{12,24}, \rOpenInt{24,36}, \rOpenInt{36,48}$, and $\rOpenInt{48,60}$. For period, it is smoothed within each of the six age groups as well as being smoothed globally. For more explicit details, we refer to either \citet{wakefield2019estimating} or \citet{li2020space}. Due to the similarity between the AP and the cluster level models, we used the AP model as the benchmark to compare the AC and APC models against to answer the question ``is it suitable to include cohort alongside age and period for subnational estimates of U5MR''.

\subsection{Prior Selection and Implementation}

We fit a Bayesian hierarchical model in which prior distributions needed to be assigned to all the parameters in the model. For the urban, rural, and the two temporal slope parameters, we used a wide, normal prior $\beta_1, \dots, \beta_4 \sim \text{Normal}\pa{0, 1000}$. For the temporal curvature terms, we used a RW2 model, $\nu_{\tilde{a}} | \tau_\nu \sim \text{RW2}\pa{\tau_\nu}$, $\eta_p | \tau_\eta \sim \text{RW2}\pa{\tau_\eta}$ and $\xi_c | \tau_\xi \sim \text{RW2}\pa{\tau_\xi}$ where $\tau_\nu$, $\tau_\eta$ and $\tau_\xi$ are the precision parameters for the age, period, and cohort curvature, respectively.

In the spatial term, $S_{r\bra{s_k}}$, there is a structured and unstructured spatial component. Independently, the structured part was modelled using an intrinsic conditional autoregressive (ICAR) model \citep{besag1991bayesian}, and the unstructured part was modelled with an independent identically distributed model. We chose to model them together using a BYM2 parameterisation \citep{riebler2016intuitive}, $S_{r\bra{s_k}} | \tau_s, \phi \sim \text{BYM2}\pa{\tau_S, \phi}$, where $\phi \in \bra{0,1}$ is the mixing parameter that measures the proportion of the marginal variance, $\tau_S^{-1}$,  that is explained by the structured spatial effect. Since we believed the period trend will be different from region-to-region whilst also being structured in space, we assumed a Type IV space-period interaction for $\delta_{p,r\bra{s_k}}$ \citep{knorr2000bayesian}. As with the main period and spatial effects, we assumed a RW2 and ICAR prior distribution, respectfully. 

To complete the prior specification, for all distributions we used penalised complexity (PC) priors \citep{simpson2017penalising} on precision and correlation components such as $\tau$ and $\phi$. A PC prior for a given model component is specified through $\text{P}\pa{\text{model parameters } > U} = p$ where $U$ is an appropriate upper bound and $p$ is the probability of the model parameter being in the upper bound. We followed \cite{li2020space} for the PC priors hyperparameters specification for each model term. The precisions for all temporal curvatures have $U = 1$ and $p = 0.01$. For the BYM2 term, the variance and scale parameters had $U = 1$ and $p = 0.01$ and $U = 1/2$ and $p = 2/3$, respectively. For the space-period interaction term $U = 1/2$ and $p = 2/3$.

We fit the spatio-temporal APC model with integrated nested Laplace approximations (INLA) as implemented in the \inla \ package \citep{rue2009approximate}. INLA provides accurate approximations of the marginal posterior distribution for random effects and hyper-parameters whilst avoiding the need for costly and time-consuming Markov-chain Monte Carlo (MCMC) sampling.

The age, period, and cohort random effects (curvatures) were fit in \inla \ using the following argument specification \texttt{f(..., model = `rw2', constr = TRUE, rankdef = 2)} to ensure the temporal linear trends included in the model are fully identifiable. Additionally, we used the \texttt{values = c(0, 6, 17.5, 29.5, 41.5, 53.5)} argument for age to ensure the non-constant age groups were interpreted correctly. The spatial random effect was fit using \texttt{model = `bym2'} and the standard adjacency-matrix. 

We fit the Type IV space-period interaction using the \texttt{model = `generic0'} argument and explicitly specified the precision matrix and associated constraints to ensure identifiability. We defined the space-period precision matrix as $\boldsymbol{Q}_\text{space-period} = \boldsymbol{Q}_\text{period} \otimes \boldsymbol{Q}_\text{space}$, where the temporal and spatial precision matrices were defined using a RW2 and ICAR structure, respectively, an $\otimes$ is the Kronecker product. To satisfy the linear constraints, $\boldsymbol{A}\boldsymbol{x} = \boldsymbol{e}$ \citep{rue2005gaussian}, the additional constraints are defined via the eigenvectors that correspond to the zero eigenvalues of $\boldsymbol{Q}_\text{space-period}$. More details on prior specification can be found in the Supplementary Material.

\subsection{Constructing the U5MR posterior}

We took independent samples from the approximated posterior distribution of the monthly hazards using \inla, but were not able to sample from the U5MR posterior directly as the formula for U5MR requires non-linear combinations of samples from the monthly hazards posterior. To generate a sample for the U5MR posterior, we first sampled from the monthly hazards joint posterior using \inla, then using the formula in Equation \eqref{Eq: U5MR General Formula}, we defined a sample from the U5MR posterior.

Using one independent sample from the posterior distribution of the monthly hazard from \inla, the $n$-th sample from the U5MR posterior for period $p$ and region $r$ is defined
\begin{equation}
    \label{Eq: U5MR General Formula}
    \begin{split}
        \text{U5MR}^{\pa{n}}_{p,r} & = 1 - \prod_{i = 1}^6 \bra{1 - \expit\pa{\logit\bra{\pi^{\pa{n}}_{\tilde{a}\bra{i},p,c,r}}} }^{z\bra{i}} \\
        & = 1 - \prod_{i = 1}^6 \bra{\frac{1}{1+\text{exp}\pa{\logit\bra{\pi^{\pa{n}}_{\tilde{a}\bra{i},p,c,r}}}}}^{z\bra{i}}        
    \end{split}
\end{equation}
where $\tilde{a} = 0.5, 6, 17.5, 29.5, 41.5, 52.5$ and $z = 1, 11, 12, 12, 12, 12$ are the midpoints and number of grouped months in each of the six discrete age groups, respectively. This formula is an adaption of the U5MR formula, Equation \eqref{Eq: Simplified U5MR Hazard Probability}, discussed previously.

As we are fitting a stratified cluster level model, we cannot directly find the regional U5MR. Instead, we find an U5MR estimate for the urban/rural stratified region and then aggregate over the urban/rural stratum to find the regional U5MR \citep{paige2022design}. The U5MR for the urban and rural areas in period $p$ for region $r$ are
\begin{align*}
    \text{U5MR}^{\pa{n}}_{p,r,\text{rural}} & = 1 - \prod_{a = 1}^{6} \bra{\frac{1}{1+\text{exp}\pa{ \beta^{\pa{n}}_1 + {t_{1,r}}{\beta^{\pa{n}}_3} + {t_{2,r}}{\beta^{\pa{n}}_4} + {\nu^{\pa{n}}_{\tilde{a}}} + {\eta^{\pa{n}}_p} + {\xi^{\pa{n}}_c} + S^{\pa{n}}_r + \delta^{\pa{n}}_{p,r}}}}^{z\bra{a}}, \\ 
    \text{U5MR}^{\pa{n}}_{p,r,\text{urban}} & = 1 - \prod_{a = 1}^{6} \bra{\frac{1}{1+\text{exp}\pa{\beta^{\pa{n}}_1 + \beta^{\pa{n}}_2 + {t_{1,r}}{\beta^{\pa{n}}_3} + {t_{2,r}}{\beta^{\pa{n}}_4} + {\nu^{\pa{n}}_{\tilde{a}}} + {\eta^{\pa{n}}_p} + {\xi^{\pa{n}}_c} + S^{\pa{n}}_r + \delta^{\pa{n}}_{p,r}}}}^{z\bra{a}}.
\end{align*}
The aggregated U5MR for period $p$ and region $r$ is 
\begin{equation}
    \label{Eq: Aggregated Region-Period U5MR}
    \text{U5MR}^{\pa{n}}_{p,r} = \text{U5MR}^{\pa{n}}_{p,r,\text{rural}} \times {q_{p,r}} +  \text{U5MR}^{\pa{n}}_{p,r,\text{urban}} \times \pa{1 - {q_{p,r}}}
\end{equation}
where $q_{p,r}$ is the proportion of the target population in period $p$ and region $r$ that is rural. We used the proportions for each period-region combination to account for both the urbanisation over the study period and the over and under sampling of urban or rural areas. A estimate for the national U5MR was defined by multiplying $\text{U5MR}_{p,r}$ by the yearly national proportions of each region \citep{wu2022modeling}. For a full distribution of the stratified, subnational (by region), and national U5MR, the above process is repeated $N = 1000$ times.

As part of our analysis, we generated predictions of U5MR for up to five years in the future. First, we defined the month-by-month periods over the prediction time-frame (January 2014 to December 2018). From the month-by-month periods, we subtracted the monthly ages (0 -- 59) to define the month-by-month (birth) cohorts. We then grouped age into the six discrete age intervals and grouped period and cohort into the yearly intervals. This process replicates the real-birth cohorts we would see in the data. Unless stated otherwise, we use the median and 95\% Credible Interval (CI) as point estimates and measure of uncertainty in the resulting analysis.

\section{Predictive validation}
\label{Section: Validation}

To evaluate the predictive ability of all the models (AP, AC, and APC), we performed a Cross-Validation (CV). The 2014 birth summaries in the 2014 KDHS are incomplete (due to 2014 being the year the survey was conducted). Consequently, we performed the CV using data from the years 2006 -- 2013 where we systematically left out 2013 for each of Kenya's 47 regions, fit the models to the remaining data, and predicted the missing 2013 observations for the given region. 

The U5MR posterior described in Section \ref{Section: Methods} will contain sampling variability from \inla, but does not propagate uncertainty arising from the (complex) survey design. To account for this and make the model estimates more comparable to the gold-standard direct estimates, we added noise from the U5MR sampling distribution to our estimates of U5MR. In other words,
\begin{equation}
    \label{Eq: U5MR Sampling Distribution}
    \widetilde{Y}^{\pa{m}}_{r} = \widehat{Y}^{\pa{m}}_{r} + \text{Normal}\pa{0, \hat{V}^{\text{Des}}_{\text{r}}}
\end{equation}
where $\widehat{Y}^{\pa{m}}_{r} = \logit \pa{ \text{U5MR}^{\pa{m}}_{r} }$ is the $m\text{th}$ draw from the estimated U5MR posterior distribution from \inla \ produced when fitting the model to a dataset missing the observations of region $r$ in the year 2013, and $\hat{V}^\text{Des}_\text{r}$ is the complex design variance for region $r$ in the year 2013 estimated using the direct estimates found using the \summer \ package. We used this U5MR normal sampling distribution since it has been shown to perform well in the context of small area estimation for complex survey data \citep{mercer2015small}.

\subsection{Assessment criteria}

To assess the predictive performance of the models, we computed the Mean Absolute Error (MAE), variance, and Mean Squared Error (MSE) against the direct estimates. Using the $N = 1000$ draws of the U5MR posterior distribution, the scores are calculated as
\begin{align*}
    \text{MAE} & = \frac{1}{R}\frac{1}{N}\sum_{r = 1}^R\sum_{m = 1}^N \abs{\widetilde{Y}^{\pa{n}}_r - Y_r} \\ 
    \text{MSE} & =  \frac{1}{R}\frac{1}{N}\sum_{r = 1}^R\sum_{m = 1}^N \bra{\widetilde{Y}^{\pa{n}}_r - Y_r}^2 \\ 
\end{align*}
where $Y_r$ is the direct estimate of the U5MR for region $r$ in 2013 on the logit scale.

In addition, we assessed the entire predictive distribution using the 95\% Interval Score \citep[IS;][]{gneiting2007strictly} and coverage. The IS is a scoring rule that transforms interval width and empirical coverage into a single score. Let $Y_r$ be the direct estimate of logit-U5MR for region $r$ and $\bra{l_{r}, u_{r}}$ be the lower and upper predictive quantiles from the U5MR sampling distribution defined in Equation \eqref{Eq: U5MR Sampling Distribution}. We used the direct estimate as they are design consistent, and at a national level are the gold standard estimate. The IS for $\alpha \in \pa{0, 1}$ is defined
\begin{equation*}
    \label{Eq: Interval Score}
    \text{IS}_{\alpha}\pa{Y} = \frac{1}{R} \sum_{r = 1}^{R} \left[ \pa{u_{r} - l_{r}} + \frac{2}{\alpha}\pa{l_{r} -Y_r}\II\pa{Y_r < l_{r}} + \frac{2}{\alpha}\pa{Y_r - u_{r}}\II\pa{Y_r > u_{r}} \right]
\end{equation*}
where $\II\pa{\cdot}$ is an indicator function that penalises $\pa{Y}_r$ falling outside the interval. When using the coverage and IS score together, the better fitting models have a (relatively) high coverage and low IS score which reflects a well-fitting model with narrow intervals that still contain the direct estimates. In contrast, a (relatively) large coverage with a high IS score indicates a poorly fitting model with wide intervals that captures all the direct estimates.

\subsection{Results}

\begin{table}[!h]
    \centering
    \caption{Validation of the sampling distribution. Averaged over all regions are the score for the Mean Absolute Error, Mean Squared Error, and Interval Score. For both the Interval Score and Coverage, we included the $\alpha = 0.50$ and $\alpha = 0.05$ level results for each model. The best score of each category is indicated with \textbf{\textit{emphasised}} text.}
    \label{Tab: Validation Scores}
    \resizebox{\textwidth}{!}{%
    \begin{tabular}{c|cc|cc|cc}
        \hline
        \multirow{2}{*}{\textbf{Model}} & \multirow{2}{*}{\textbf{\begin{tabular}[c]{@{}c@{}}Mean Absolute Error\\ ($\bs{\times 10^{-2}}$)\end{tabular}}} & \multirow{2}{*}{\textbf{\begin{tabular}[c]{@{}c@{}}Mean Squared Error\\ ($\bs{\times 10^{-2}}$)\end{tabular}}} & \multicolumn{2}{c|}{$\bs{\alpha = 0.50}$} & \multicolumn{2}{c}{\textbf{$\bs{\alpha = 0.05}$}} \\
        &  &  & \textbf{Interval Score} & \textbf{Coverage (\%)} & \textbf{Interval Score} & \textbf{Coverage (\%)} \\ \hline
        Age-Period & 70.1 & 77.4 & 72.3 & 44.2 & 145.5 & \textbf{\textit{88.4}} \\ 
        Age-Cohort & 71.3 & 80.2 & 73.9 & \textbf{\textit{48.8}} & 155.0 & \textbf{\textit{88.4}} \\ 
        Age-Period-Cohort & \textit{\textbf{69.0}} & \textit{\textbf{73.9}} & \textit{\textbf{71.6}} & 44.2 & \textbf{\textit{142.6}} & \textbf{\textit{88.4}} \\ \hline
    \end{tabular}%
    }
\end{table}

Table \ref{Tab: Validation Scores} shows the validation scores of the sampling distribution for each of the AP, AC and APC models. For the MAE and MSE, the APC model performed the best though it is only marginally better than the AP model. At the level of $\alpha = 0.5$, the APC model had the best IS score (closely matched by the AP model) and the AC model had the best coverage. The large coverage of the AC model does not reflect a better fitting model in comparison to the AP and APC models since as the larger IS implies the better coverage comes from wider intervals rather than being better fitting. For the IS score at a level of $\alpha = 0.05$, the APC model was the best and all three models had an equal coverage. 

For each of the scores, the true values were taken as the direct estimates. In situations where data sparsity is of no concern, this is acceptable as the direct estimates are design consistent. However, for sparse data situations, i.e., when modelling subnational estimates of U5MR using survey data from LMIC, the direct estimates suffer excess variability. To better discern between models, we included the IS and coverage at the 50\% CI as well as at 95\%. Considering all model scores, the AP and APC models clearly outperformed the AC model. As to which of the AP and APC models are better, the difference is marginal. Given this, the choice to use an AP or APC model in the context of modelling Kenyan U5MR is not about model performance, but rather about the researchers wants. This is due to models performing equally well. If the researcher wishes to explore the influence of cohort within the context of Kenyan U5MR, then an APC model is appropriate and will produce results in line with the AP model (and hence cluster level models) and allow the desired analysis.

\begin{figure}[!h]
    \centering
    \caption{Ridge plot of the sampling distribution for each of the predictions for 2013. On the bottom of each plot is the direct estimates for 2013 indicated by a red cross.}
    \label{Fig: Kenya_adm1_u5_cv_ridgeplot}
    \includegraphics[width=0.7\textwidth]{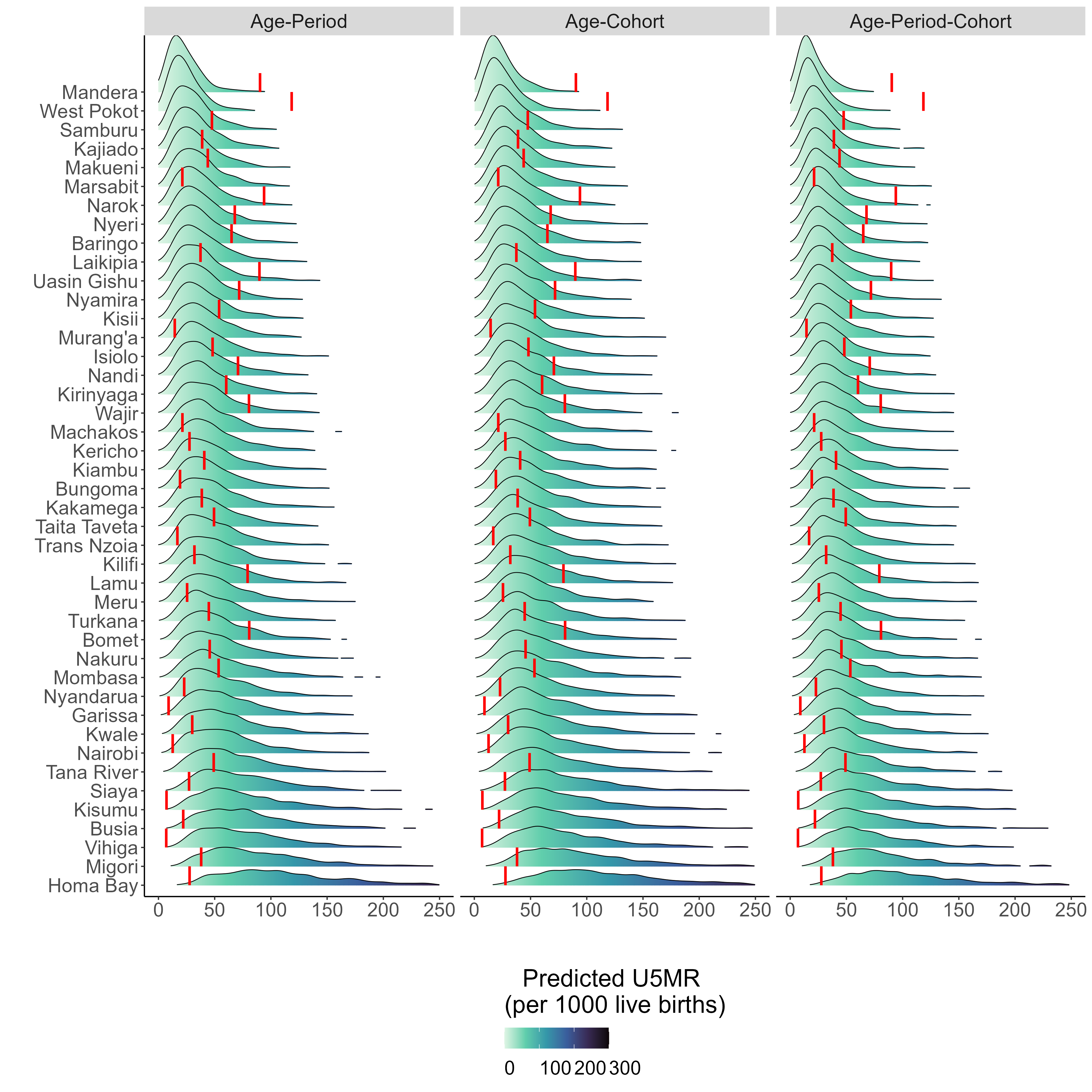}
\end{figure}

Figure \ref{Fig: Kenya_adm1_u5_cv_ridgeplot} shows a ridge plot of the sampling distribution for each region for the year 2013 produced during the CV process. For each ridge, there is a vertical red line that indicates the direct estimate for that region. During 2013, there for four regions (Elgeyo-Marakwet, Embu, Kitui, and Tharaka-Nithi) that did not have any known under-five deaths. Consequently, these regions do not have a direct estimate, and we did not include their ridge. Across all three facets, the red lines being within most of the ridges 95\% CI gives a visual representation of each models Coverage. The direct estimate for regions such as West Pokot, Marsabit, Vihiga, and Homa Bay clearly all lie outside the 95\% CI. The APC model had a marginally better IS, and this can be seen when considering the width of the mass of the APC models distribution which is slightly narrower than those of the AP and AC models. Finally, when considering the posterior uncertainty (namely the right-hand tail of each distribution), those from the APC model are narrower than those of the AP and AC models. To see this, consider the red line of the direct estimate in comparison to the end of the right-hand tail such as those from the regions of Madera and West Pokot.

\section{Kenyan Under-Five Mortality Rate}

In the following section, we used the 2014 KDHS years 2006 -- 2013 (excluding the incomplete 2014 year) for estimation, and the years 2014 -- 2018 for prediction. The main goal of this paper is producing subnational estimates and predictions of U5MR. Before showing the subnational results, we first show the results for a second validation that was performed. The second validated compared the national direct and national FH model results to the results of the subnational AP, AC and APC models proportionally aggregated to the national level. At the national level, data sparsity is less of a concern than at the subnational level, and the direct estimates do not suffer the large variability. Consequently, we do not need to propagate uncertainty into the results for the AP, AC and APC models below like in the subnational validation in Section \ref{Section: Validation}.  After the validation at the national level, we present the main subnational results from the APC model along with additional APC specific analysis.

\subsection{National results}

\begin{figure}[!h]
    \centering
    \caption{National estimates and prediction of Under-Five Mortality Rate for Kenya between the years 2006 -- 2018. The years 2006--2013 are estimates from the data and the years 2014--2018 are predictions. The red, blue, green, yellow and purple lines indicate results from the direct, Fay-Herriot, Age-Period, Age-Cohort and Age-Period-Cohort models, respectively. The horizontal dotted black line indicate the World Health Organisations Sustainable Development Goal 3.2.}
    \label{Fig: Kenya_natl_error_plot}
    \includegraphics[width=0.7\textwidth]{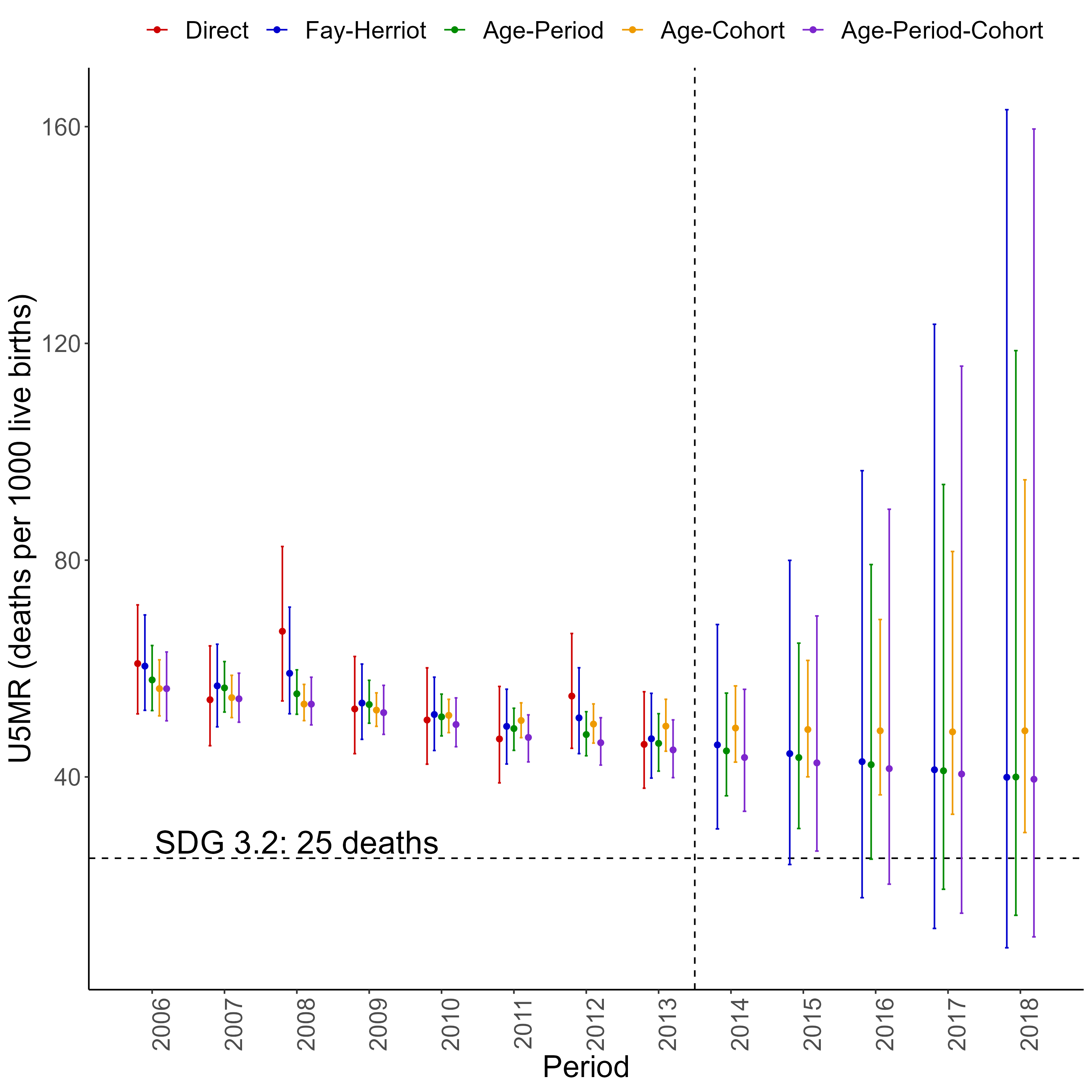}
\end{figure}

The focus of this paper is subnational estimates of U5MR using a spatio-temporal APC model. However, we compared the national \citep{horvitz1952generalization} and FH estimates \citep{fay1979estimates} to those of the subnational AP, AC and APC models where the estimates were aggregated using the regional-urban/rural and country-regional proportions . The direct estimates are design consistent, but when data is sparse, they have a large amount of uncertainty despite being the gold standard at the national level. Consequently, it is advisable to compare against national direct estimates as data is plentiful. The FH estimates are a transformation of the direct estimates modelled smoother across time (not space). The direct and FH estimates were found using the \summer \ package of \citet{li2020space}, and for the AP, AC and APC models, the samples from the U5MR posterior were proportionally aggregated to the national level.

Figure \ref{Fig: Kenya_natl_error_plot} shows the national estimates and predictions from each method along with their associated uncertainty over the years 2006 --- 2018. The vertical dotted line indicates where the predictions begin. The direct estimates (red) cannot be used for prediction whereas the FH (blue), AP (green), AC (yellow) and APC (purple) models can. Consequently, there are no direct estimates past 2013. The horizontal dotted line indicates the WHOs SDG 3.2, the target of 25 deaths per 1000 live births. 

Overall, the AP and APC models follow similar downwards trends and converging to one another over the prediction years. The AC model is extremely similar to the APC model for the first two years, and then stays relatively flat for the remaining (estimation and prediction) years. As expected, the CI for the direct estimates is always larger than those of the FH, AP, AC and APC models. For all the estimation years, the AP, AC and APC models medians all lie within the CI for both the direct and FH models. Furthermore, for all estimation years except 2008 and 2012, the CI for the AP, AC and APC models lie within the CI for the direct estimates. For the prediction years, the medians and CIs of the AP, AC and APC models lie within the CI for the FH model. The difference between the AC and the other model estimates becomes more pronounced the further into the predictions, with the AC model medians having a flatter curve and narrower CIs. This may be due to AC model not being as influenced by yearly fluctuations \citep{best2018premature}, that would be captured in the FH, AP and APC models, as it predicts using the less volatile cohort trend.

\subsection{Subnational results}

Table \ref{Tab: Model Summaries} show the posterior summaries for each of the AP, AC and APC models. The interpretation for each of the parameters shared between all models is consistent. For example: zero is in the CI for each of the rural and intercepts as well as for the age slope parameter; the quantiles for each of the overdispersion $d$, age precision $\tau_{\nu}$, and spatial precision $\tau_S$ parameters are all similar to one another; the spatial mixing parameter $\phi$ for each indicates the residual variation described by structured and unstructured is around 0.5; and finally, the spatio-temporal precision $\tau_\delta$ varies between the three, but the medians are all within one another’s CI. 

\begin{table}[!t]
    \centering
    \caption{Posterior quantile summaries for each of the Age-Period, Age-Cohort and Age-Period-Cohort models.}
    \label{Tab: Model Summaries}
    \resizebox{\textwidth}{!}{%
    \begin{tabular}{c|ccc|ccc|ccc}
        \hline
        \multirow{2}{*}{\textbf{Parameter}} & \multicolumn{3}{c|}{\textbf{Age-Period}} & \multicolumn{3}{c|}{\textbf{Age-Cohort}} & \multicolumn{3}{c}{\textbf{Age-Period-Cohort}} \\
         & \textbf{Lower} & \textbf{Median} & \textbf{Upper} & \textbf{Lower} & \textbf{Median} & \textbf{Upper} & \textbf{Lower} & \textbf{Median} & \textbf{Upper} \\ \hline
        Rural intercept & -42.4 & -0.5 & 41.5 & -42.2 & -0.5 & 41.2 & -42.2 & -0.3 & 41.7 \\
        Urban intercept & -42.4 & -0.4 & 41.5 & -42.2 & -0.5 & 41.2 & -42.2 & -0.2 & 41.7 \\
        Age slope & -3.2 & -0.2 & 2.9 & -3.0 & -0.2 & 2.6 & -3.1 & -0.1 & 3.0 \\
        Period slope & -11.0 & -0.4 & 10.3 & - & - & - & -11.4 & -0.8 & 9.9 \\
        Cohort slope & - & - & - & -7.3 & -0.2 & 6.9 & - & - & - \\ \hline
        Betabinomial overdispersion ($\times 10^{-5}$), $d$ & 2.7 & 22.1 & 90.6 & 2.8 & 22.8 & 91.2 & 2.0 & 18.0 & 81.0 \\
        Age precision, $\tau_{\nu}$ & 0.4 & 1.0 & 2.2 & 0.4 & 1.0 & 2.2 & 0.4 & 0.9 & 2.1 \\
        Period precision, $\tau_{\eta}$ & 13.6 & 386.6 & 29893.9 & - & - & - & 8.3 & 130.0 & 3107.6 \\
        Cohort precision, $\tau_{\xi}$ & - & - & - & 15.6 & 420.8 & 37392.4 & 7.1 & 92.8 & 1862.8 \\
        Region precision, $\tau_{S}$ & 4.9 & 8.4 & 13.8 & 4.9 & 8.3 & 13.8 & 4.9 & 8.3 & 13.7 \\
        Region mixing, $\phi$ & 0.1 & 0.5 & 0.9 & 0.1 & 0.5 & 0.9 & 0.1 & 0.5 & 0.9 \\
        Space-period precision, $\tau_{\delta}$ & 111.3 & 896.1 & 20630.6 & 99.2 & 645.5 & 10503.5 & 90.7 & 746.2 & 13834.2 \\ \hline
    \end{tabular}%
    }
\end{table}

\begin{figure}[!b]
    \centering
    \caption{Subnational estimates of Under-Five Mortality Rate for Kenya from the Age-Period-Cohort model. The years 2006--2013 are estimates from the data and the years 2014--2018 are predictions. Figure (a) shows the estimate per 1000 live births and Figure (b) shows the width of the 95\% Credible Interval.}
    \label{Fig: Subnational Maps}
    \begin{subfigure}[b]{0.5\textwidth}
        \centering
        \includegraphics[width=\textwidth]{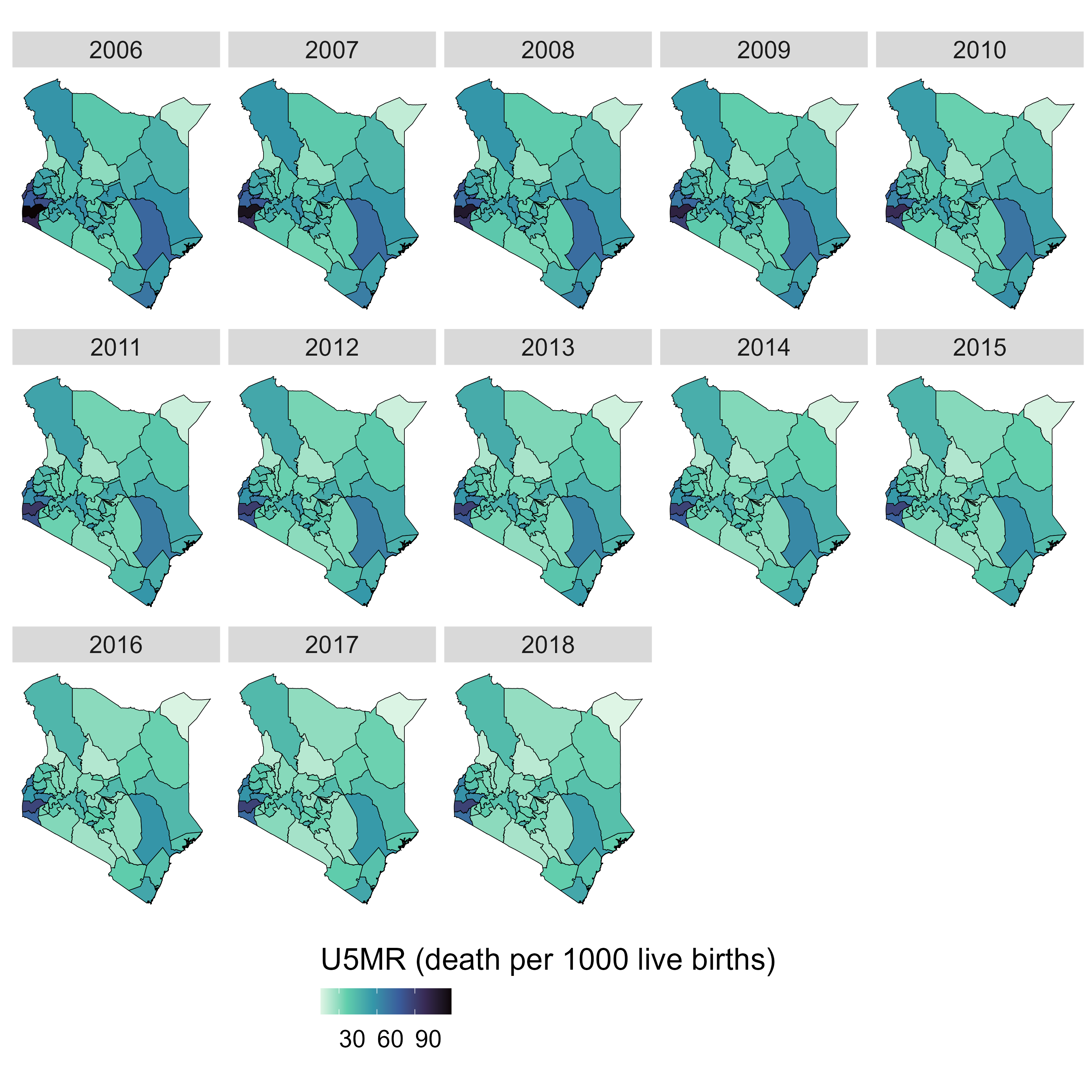}
        \caption{Subnational Under Five Mortality Rate.}
        \label{Fig: Kenya_apc_adm1_u5_estimate_plot}
    \end{subfigure}%
    \begin{subfigure}[b]{0.5\textwidth}
        \centering
        \includegraphics[width=\textwidth]{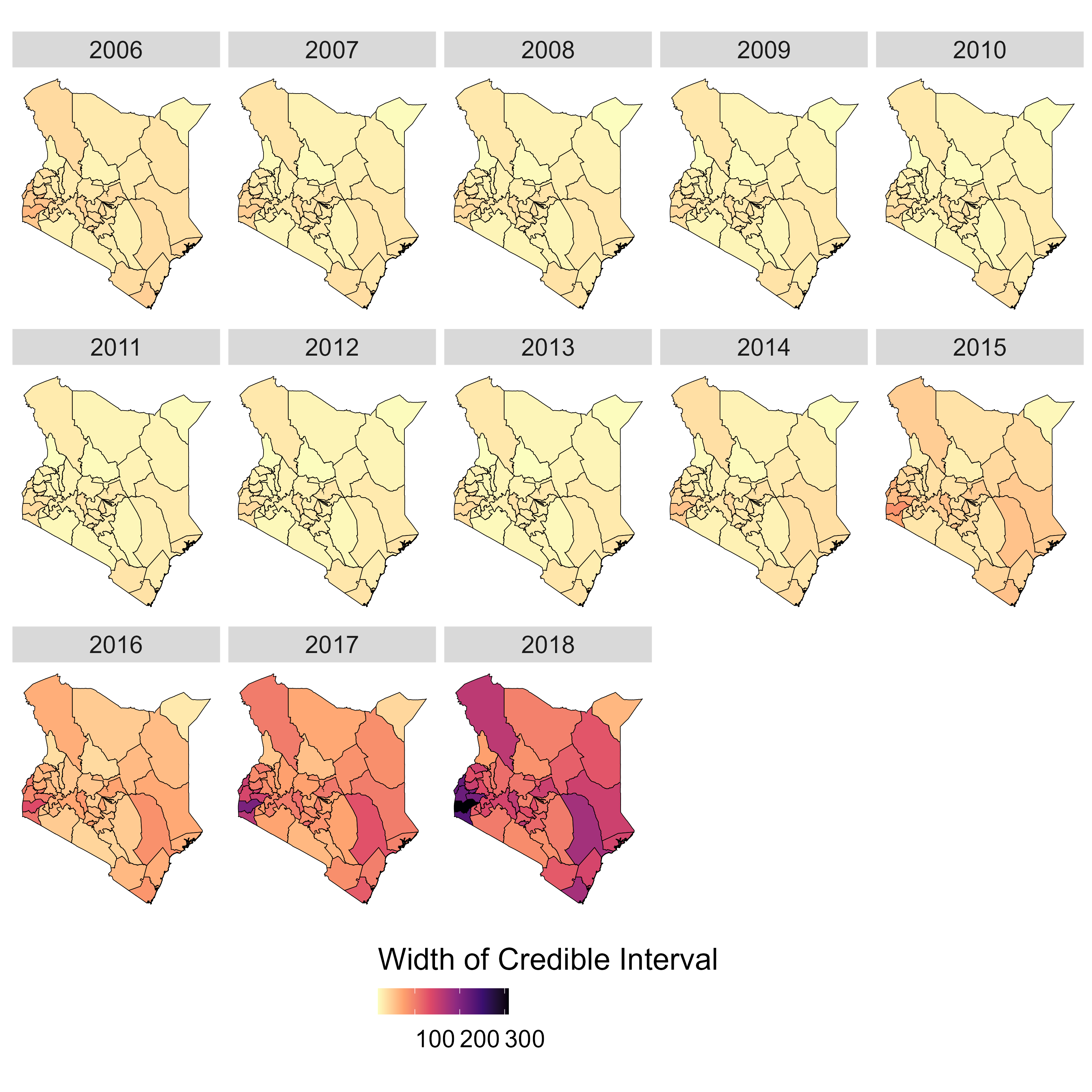}
        \caption{Width of 95\% credible interval.}
        \label{Fig: Kenya_apc_adm1_u5_width_plot}
    \end{subfigure}
\end{figure} 

Figure \ref{Fig: Subnational Maps} shows the subnational maps of Kenya for each year with Figure \ref{Fig: Kenya_apc_adm1_u5_estimate_plot} being the estimated values of U5MR per 1000 live births and Figure \ref{Fig: Kenya_apc_adm1_u5_width_plot} being the width of the 95\% CI. The spatial structure is seen by neighbouring regions having similar estimates (i.e., regions in the west having higher estimates in comparison to the rest of Kenya). When comparing the estimate and width map, regions with a higher U5MR have a larger width. This is due to the beta-binomial likelihood and associated mean-variance relationship. Overall, both the U5MR and CI width are reducing year-on-year from years 2006 -- 2013. For the forecasted years, the estimate of U5MR stays consistent whereas the CI gets notably larger the further into the future the prediction is.

\begin{figure}[!t]
    \centering
    \caption{Estimate of Under Five Mortality Rate per 1000 live births from the Age-Period-Cohort model for each of the 47 counties in Kenya. The vertical dashed line indicates where the forecast starts.}
    \label{Fig: Kenya_apc_adm1_u5_lineplot}
    \includegraphics[width=0.7\textwidth]{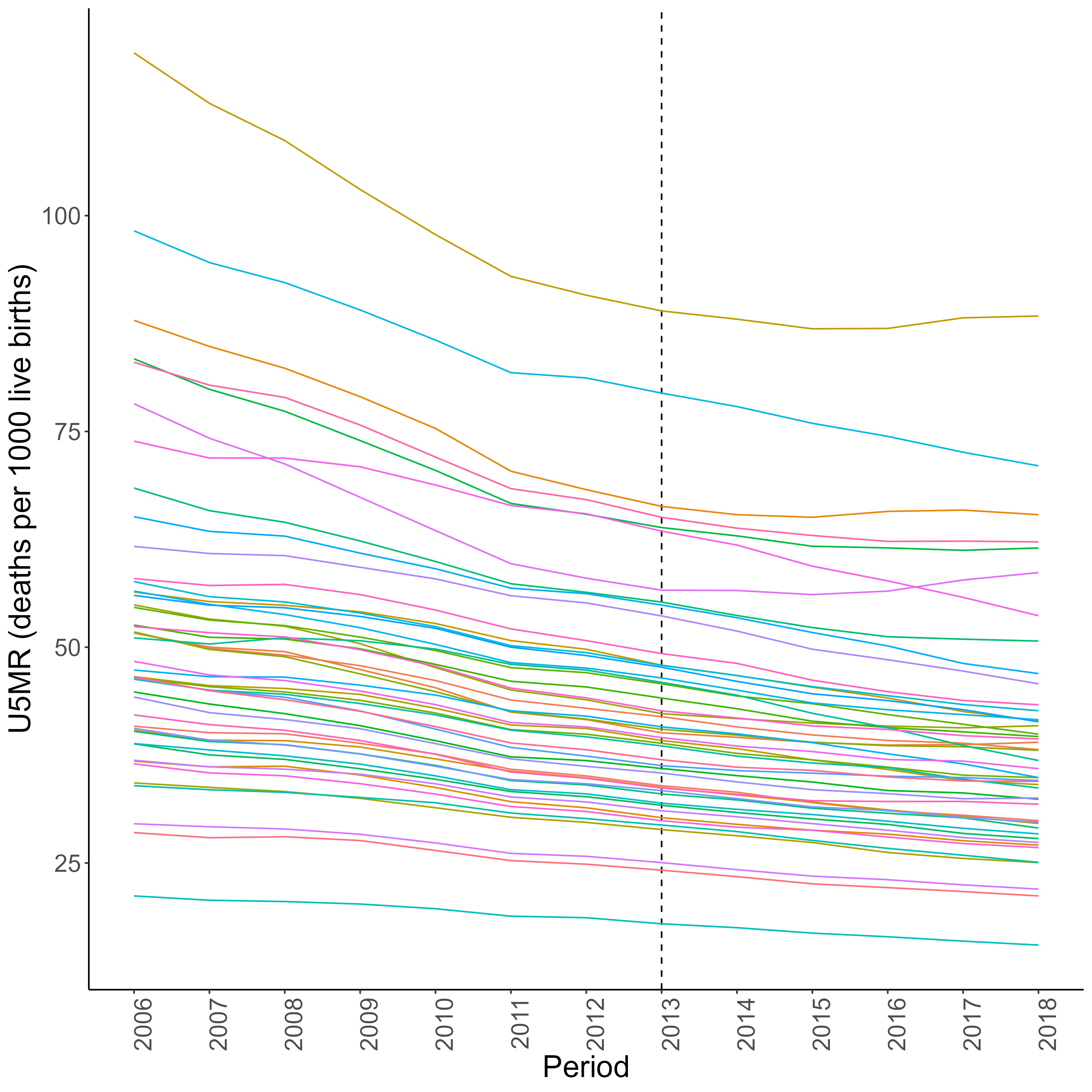}
\end{figure}

Figure \ref{Fig: Kenya_apc_adm1_u5_lineplot} shows the region-specific U5MR per 1000 live births. Almost all the regions show a declining trend in U5MR from 2006 -- 2013. For the years 2014 -- 2018, the U5MR for many regions either plateaus or declines very slightly. The overlapping in each of the lines is indicative of a space-period interaction; if all the lines were parallel (on the logit scale), there is not space-period interaction. Across all regions in Kenya, there is a general downwards trend in U5MR from the years 2006 -- 2013. For the predicted years 2014 -- 2018, this trend is shown to continue but with a gradual decline in the rate of reduction. This is reflected in the national plot.  

\begin{figure}[!t]
    \centering
    \caption{Age specific Under Five Mortality Rate by cohort and year for Kenya on the logit scale. The years 2006--2013 are estimates from the data and the years 2014--2018 are predictions.}
    \label{Fig: Age-specific U5MR}
    \begin{subfigure}[b]{0.5\textwidth}
        \centering
        \includegraphics[width=\textwidth]{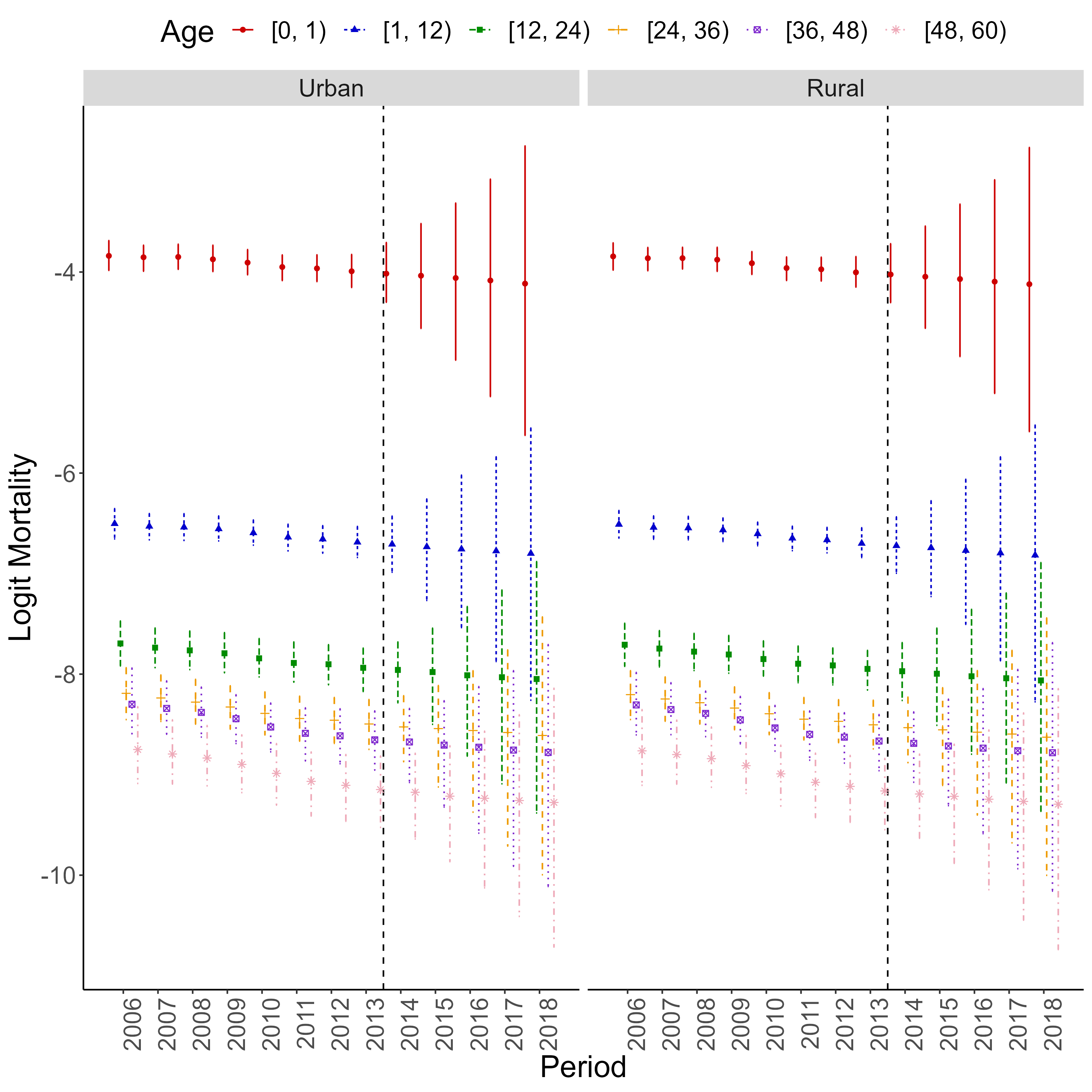}
        \caption{By period and urban/rural stratification.}
        \label{Fig: Kenya_apc_adm1_u5_agePeriodStrata_lineplot}
    \end{subfigure}%
    \begin{subfigure}[b]{0.5\textwidth}
        \centering
        \includegraphics[width=\textwidth]{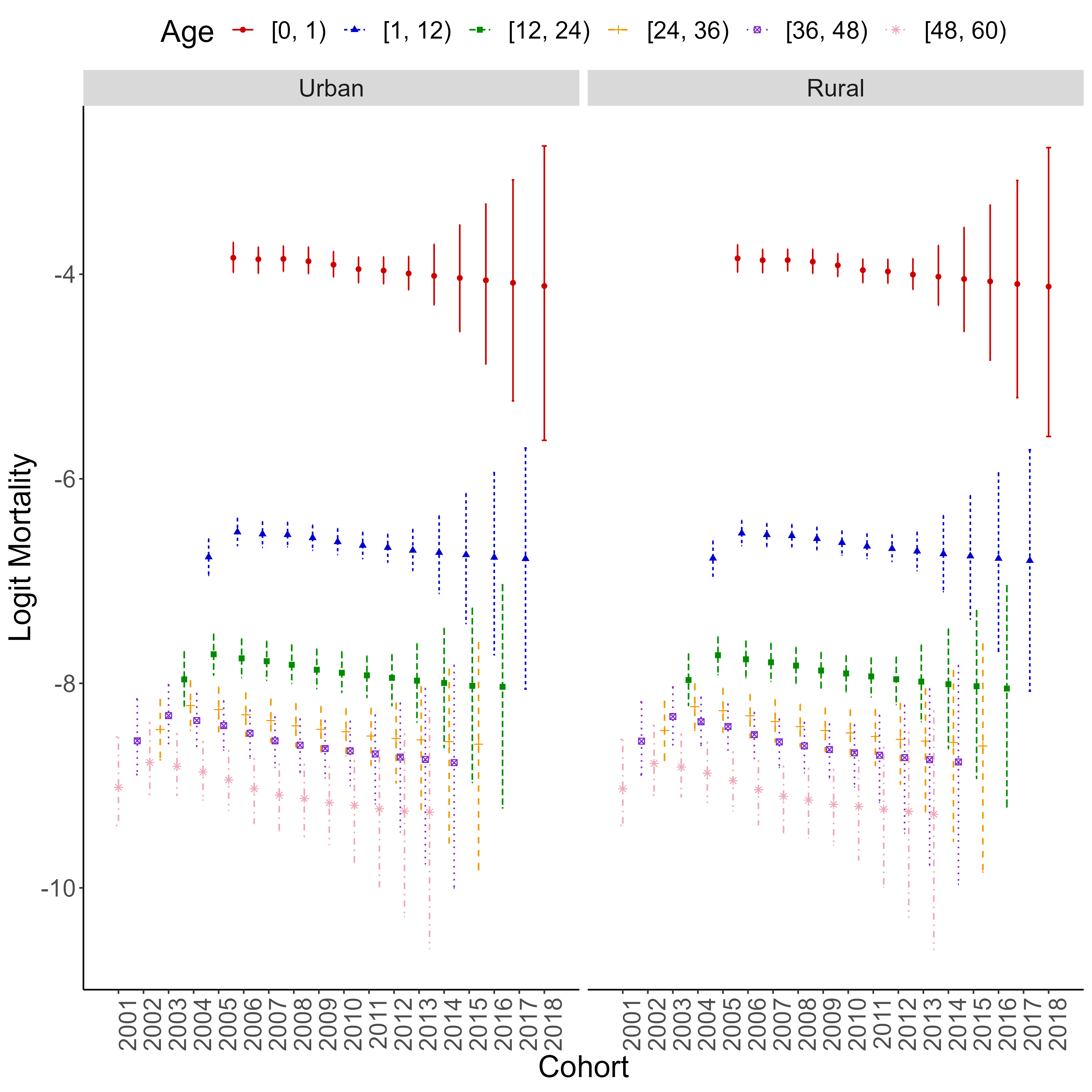}
        \caption{By cohort and urban/rural stratification.}
        \label{Fig: Kenya_apc_adm1_u5_ageCohortStrata_lineplot}
    \end{subfigure}
\end{figure}

Finally, Figure \ref{Fig: Age-specific U5MR} shows the age-specific estimated and predicted mortality by period, Figure \ref{Fig: Kenya_apc_adm1_u5_agePeriodStrata_lineplot}, and by birth cohort, Figure \ref{Fig: Kenya_apc_adm1_u5_ageCohortStrata_lineplot}. Due to most deaths coming in the first month of life, we plotted these on the logit scale to better compare between the different age-groups. Furthermore, we include a minor within-period and within-cohort jitter to better discern between the six age groups as all but the $\rOpenInt{0,1}$ and $\rOpenInt{1, 12}$ groups become increasing similar. 

For each age group in Figure \ref{Fig: Kenya_apc_adm1_u5_agePeriodStrata_lineplot}, the mortality is reducing from period-to-period. The reduction is more noticeable in the older age groups with the younger ages appearing flatter. For each age group, there is an increase in the mortality in the rural stratum when compared to the urban stratum. When considering all age groups in a given year, there is a clear decrease in mortality in the older age groups. The most noticeable difference is between ages $\rOpenInt{0, 1}$ and $\rOpenInt{1, 11}$. Additionally, the oldest three age groups are much closer to one another than they are to the youngest age groups.

For increasing age, Figure \ref{Fig: Kenya_apc_adm1_u5_ageCohortStrata_lineplot} shows the age-specific cohort mortality is reducing. Similarly to the age-specific period mortality, the difference between the $\rOpenInt{0, 1}$ and $\rOpenInt{1, 11}$ age groups is the largest and the oldest three age groups are the most similar. Furthermore, there is an increase in mortality for each age-group in the rural stratum compared to the urban stratum. As these estimates of mortality are representative of the real-world, not all age-cohort combinations are present. For example, there is no 2001 cohort mortality for ages $\rOpenInt{0, 1}$ since we only use data that starts in 2006, and the youngest cohort for this age is 2006. For the first to second cohort in all but the $\rOpenInt{0, 1}$ age group, there appears to be an uptick. This could be attributed to under-reporting due to accuracy in the response. For the earlier cohorts, respondents will need to know exact birth and death dates over a larger time frame which may lead to inaccuracies.  

Figure \ref{Fig: Age-specific U5MR}, and in particular Figure \ref{Fig: Kenya_apc_adm1_u5_ageCohortStrata_lineplot}, highlight the additional insight and interpretation opportunities available when using an APC framework to estimate and predict subnational U5MR. Further results are presented in the Supplementary Material.

\section{Conclusions}

In this paper, we fitted a spatio-temporal APC model to the 2014 KDHS survey data and produced subnational estimates for U5MR. The inclusion of cohort alongside age and period in the context of U5MR is novel as previous spatio-temporal methods only use age and period temporal trends. We performed a subnational predictive validation against the gold-standard weighted (direct) estimates \citep{horvitz1952generalization}, and compared the results of the AC and APC models against those of the AP model to assess whether smoothing across estimates of (birth) cohort is suitable in in the context of subnational estimation of U5MR. Finally, we produced subnational estimates and predictions of U5MR for the years 2006--2013 and 2014--2018, respectively.

Direct estimates are the gold standard when data is plentiful since they are design consistent. However, when data is sparse, such as at the subnational level, the direct estimates suffer from a large amount of sampling variability. As an alternative to direct estimates, the Fay Herriot model smooths a possibly transformed version of the direct estimates across time, and cluster level models smooth across trends in age, period, and space \citep{wakefield2019estimating, li2020space}. The theme across the latter two methods is to smooth across trends in time and/or space to reduce the variability in the estimates of U5MR. Of the models used in practise now, none of them smooth across trends in cohort. Cohort is readily available in the DHS survey but is often overlooked with current methods favouring the more important age and period temporal trends. A possible explanation is that including cohort alongside age and period leads to problems relating to lack of identifiability of the temporal trends \citep{holford1983estimation} which are further exacerbated when the temporal data is grouped into non-equal interval widths \citep{holford2006approaches}. 

Predictions of U5MR are vital to advise on policy and interventions to meet the United Nations SDG 3.2 for each country to have a U5MR of 25 per 1000 live births by 2030. In particular, the production of estimates and predictions at a subnational level are a major objective as this is the administrative level where any intervention aimed at achieving the SDG 3.2 will be implemented. Given the importance and popularity of including cohort in analysis of mortality rates in general, it is natural to want to incorporate a cohort term to produce stable estimates and predictions of U5MR in LMICs where, for data sparsity reasons, these are vital for the effective recommendation of future policy.

As part of the analysis, we validated the APC model for subnational prediction using several well recognised metrics. The results of the validation showed the AP and APC models performed equally well and were almost indistinguishable from one another in the context of Kenyan U5MR. In contrast, the AC model was noticeable worse. Consequently, the difference between the AP and APC models and the AC model confirms the importance of including period in subnational estimates of U5MR, but the closeness in the AP and APC validation results show that the APC model is suitable, and cohort could be important when modelling Kenyan U5MR. When considering between using the APC or the AP model for Kenyan U5MR, the choice is down to the goal of the user. If the researcher wishes to explore the effect of cohort, then they do not lose anything by fitting an APC model instead of an AP model, but they would if they fit an AC model instead of AP model. Similarly, whilst this is a clear outcome for Kenya, this is not true for all LMICs with DHS data. For a given country, the motivation of including cohort should be carefully considered, as well as how the results and model performance of the APC model compare to that of those AP model. 

The aim of this paper is to produce subnational estimates for U5MR that aid in policy and intervention decisions to achieve the SDG 3.2. The current model is tested on one of the richer datasets the DHS has to offer at what is known as an Admin-1 subnational level. Since any interventions are implemented on the Admin-2 level, a finer subnational scale than the Admin-1 level, predictions at this level and for a wider range of datasets are required. In addition, the models currently used in the literature allow for multiple surveys for one country to be used.

A longer-term goal involves model selection and in particular, interaction term selection. Across the smoothing methods for U5MR, a space-period interaction is often included but there has been little research conducted into other temporal-spatial and temporal-temporal interactions. The inclusion of additional temporal-temporal interactions may raise additional (lack-of) identification issues, but if the linear trends in the model are identifiable in the first place, these interactions can be included \cite{smith2016review}. The main work of this extension is the development of a thorough selection criteria to assess if the additional insight is worth the added complexity and computational cost to include the interaction. We do note, however, that we have developed an efficient computational implementation based on INLA. 

To summarise, the novel inclusion of cohort alongside age and period when modelling U5MR is a suitable method for producing smooth subnational estimates. Immediate extensions include the use of multiple surveys for one country and application to an increased number of countries from the DHS and production of estimates on a finer subnational scale. Longer term goals include the development of a model selection procedure for additional interactions.

\section*{Data Statement}

Data from the DHS is free to download after registration with a suitable project that allows access to the Kenyan dataset specifically. The full code for implementing the APC (and any other) model considered in this paper can be found at \url{https://github.com/connorgascoigne/Subnational-U5MR-with-APC-models}.

\section*{Competing interests}
No competing interest is declared.

\section*{Acknowledgments}

The authors would like acknowledge Alana McGovern, Zehang Richard Li and Yunhan Wu for their help with producing the results for the direct estimates, Fay-Herroit estimates and subnational weights for proportional aggregation.

\clearpage

\bibliographystyle{abbrvnat}
\bibliography{library}

\begin{thebibliography}{37}
\providecommand{\natexlab}[1]{#1}
\providecommand{\url}[1]{\texttt{#1}}
\expandafter\ifx\csname urlstyle\endcsname\relax
  \providecommand{\doi}[1]{doi: #1}\else
  \providecommand{\doi}{doi: \begingroup \urlstyle{rm}\Url}\fi

\bibitem[Besag et~al.(1991)Besag, York, and Molli{\'e}]{besag1991bayesian}
J.~Besag, J.~York, and A.~Molli{\'e}.
\newblock {Bayesian image restoration, with two applications in spatial
  statistics}.
\newblock \emph{{Annals of the Institute of Statistical Mathematics}},
  43\penalty0 (1):\penalty0 1--20, 1991.

\bibitem[Best et~al.(2018)Best, Haozous, de~Gonzalez, Chernyavskiy, Freedman,
  Hartge, Thomas, Rosenberg, and Shiels]{best2018premature}
A.~F. Best, E.~A. Haozous, A.~B. de~Gonzalez, P.~Chernyavskiy, N.~D. Freedman,
  P.~Hartge, D.~Thomas, P.~S. Rosenberg, and M.~S. Shiels.
\newblock {Premature mortality projections in the USA through 2030: a modelling
  study}.
\newblock \emph{The Lancet Public Health}, 3\penalty0 (8):\penalty0 e374--e384,
  2018.

\bibitem[Chernyavskiy et~al.(2020)Chernyavskiy, Little, and
  Rosenberg]{chernyavskiy2020spatially}
P.~Chernyavskiy, M.~P. Little, and P.~S. Rosenberg.
\newblock Spatially varying age--period--cohort analysis with application to
  {US} mortality, 2002--2016.
\newblock \emph{Biostatistics}, 21\penalty0 (4):\penalty0 845--859, 2020.

\bibitem[Clark et~al.(2013)Clark, Kahn, Houle, Arteche, Collinson, Tollman, and
  Stein]{clark2013young}
S.~J. Clark, K.~Kahn, B.~Houle, A.~Arteche, M.~A. Collinson, S.~M. Tollman, and
  A.~Stein.
\newblock Young children's probability of dying before and after their mother's
  death: {A} rural {S}outh {A}frican population-based surveillance study.
\newblock \emph{PLoS Medicine}, 10\penalty0 (3):\penalty0 e1001409, 2013.

\bibitem[Clayton and Schifflers(1987)]{clayton1987b}
D.~Clayton and E.~Schifflers.
\newblock {Models for temporal variation in cancer rates. II:
  Age--period--cohort models}.
\newblock \emph{{Statistics in Medicine}}, 6\penalty0 (4):\penalty0 469--481,
  1987.

\bibitem[Ettarh and Kimani(2012)]{ettarh2012determinants}
R.~Ettarh and J.~Kimani.
\newblock {Determinants of under-five mortality in rural and urban Kenya}.
\newblock \emph{{Rural and Remote Health}}, 12\penalty0 (1):\penalty0 3--11,
  2012.

\bibitem[Fay and Herriot(1979)]{fay1979estimates}
R.~E. Fay and R.~A. Herriot.
\newblock Estimates of income for small places: {A}n application of
  {J}ames-{S}tein procedures to census data.
\newblock \emph{Journal of the American Statistical Association}, 74\penalty0
  (366a):\penalty0 269--277, 1979.

\bibitem[Gascoigne and Smith(2023)]{gascoigne2023penalized}
C.~Gascoigne and T.~Smith.
\newblock Penalized smoothing splines resolve the curvature identifiability
  problem in age-period-cohort models with unequal intervals.
\newblock \emph{{Statistics in Medicine}}, 42\penalty0 (12):\penalty0
  1888--1908, 2023.

\bibitem[Gneiting and Raftery(2007)]{gneiting2007strictly}
T.~Gneiting and A.~E. Raftery.
\newblock Strictly proper scoring rules, prediction, and estimation.
\newblock \emph{Journal of the American Statistical Association}, 102\penalty0
  (477):\penalty0 359--378, 2007.

\bibitem[Holford(1983)]{holford1983estimation}
T.~R. Holford.
\newblock The estimation of age, period and cohort effects for vital rates.
\newblock \emph{Biometrics}, 39\penalty0 (2):\penalty0 311--324, 1983.

\bibitem[Holford(2006)]{holford2006approaches}
T.~R. Holford.
\newblock Approaches to fitting age-period-cohort models with unequal
  intervals.
\newblock \emph{{Statistics in Medicine}}, 25\penalty0 (6):\penalty0 977--993,
  2006.

\bibitem[Horvitz and Thompson(1952)]{horvitz1952generalization}
D.~G. Horvitz and D.~J. Thompson.
\newblock A generalization of sampling without replacement from a finite
  universe.
\newblock \emph{Journal of the American Statistical Association}, 47\penalty0
  (260):\penalty0 663--685, 1952.

\bibitem[{Kenya National Bureau of Statistics, Ministry of Health/Kenya,
  National AIDS Control Council/Kenya, Kenya Medical Research Institute, and
  National Council for Population and Development/Kenya}(2015)]{KDHS2014}
{Kenya National Bureau of Statistics, Ministry of Health/Kenya, National AIDS
  Control Council/Kenya, Kenya Medical Research Institute, and National Council
  for Population and Development/Kenya}.
\newblock Kenya demographic and health survey 2014, 2015.
\newblock URL \url{http://dhsprogram.com/pubs/pdf/FR308/FR308.pdf}.

\bibitem[Knorr-Held(2000)]{knorr2000bayesian}
L.~Knorr-Held.
\newblock Bayesian modelling of inseparable space-time variation in disease
  risk.
\newblock \emph{{Statistics in Medicine}}, 19\penalty0 (17-18):\penalty0
  2555--2567, 2000.

\bibitem[Knorr-Held and Rainer(2001)]{knorr2001projections}
L.~Knorr-Held and E.~Rainer.
\newblock {Projections of lung cancer mortality in West Germany: A case study
  in Bayesian prediction}.
\newblock \emph{Biostatistics}, 2\penalty0 (1):\penalty0 109--129, 2001.

\bibitem[Li et~al.(2020)Li, Martin, Dong, Fuglstad, Godwin, Paige, Riebler,
  Clark, and Wakefield]{li2020space}
Z.~R. Li, B.~D. Martin, T.~Q. Dong, G.-A. Fuglstad, J.~Godwin, J.~Paige,
  A.~Riebler, S.~Clark, and J.~Wakefield.
\newblock \emph{Space-Time Smoothing of Demographic and Health Indicators using
  the R Package SUMMER}, 2020.

\bibitem[Macharia et~al.(2019)Macharia, Giorgi, Thuranira, Joseph, Sartorius,
  Snow, and Okiro]{macharia2019sub}
P.~M. Macharia, E.~Giorgi, P.~N. Thuranira, N.~K. Joseph, B.~Sartorius, R.~W.
  Snow, and E.~A. Okiro.
\newblock {Subnational variation and inequalities in under-five mortality in
  Kenya since 1965}.
\newblock \emph{BMC Public Health}, 19\penalty0 (1):\penalty0 1--12, 2019.

\bibitem[Macharia et~al.(2021)Macharia, Joseph, Sartorius, Snow, and
  Okiro]{macharia2021subnational}
P.~M. Macharia, N.~K. Joseph, B.~Sartorius, R.~W. Snow, and E.~A. Okiro.
\newblock Subnational estimates of factors associated with under-five mortality
  in kenya: a spatio-temporal analysis, 1993--2014.
\newblock \emph{BMJ Global Health}, 6\penalty0 (4):\penalty0 e004544, 2021.

\bibitem[Mart{\'\i}nez-Miranda et~al.(2016)Mart{\'\i}nez-Miranda, Nielsen, and
  Nielsen]{martinez2016simple}
M.~D. Mart{\'\i}nez-Miranda, B.~Nielsen, and J.~P. Nielsen.
\newblock {Simple benchmark for mesothelioma projection for Great Britain}.
\newblock \emph{Occupational and Environmental Medicine}, 73\penalty0
  (8):\penalty0 561--563, 2016.

\bibitem[Mercer et~al.(2015)Mercer, Wakefield, Pantazis, Lutambi, Masanja, and
  Clark]{mercer2015small}
L.~Mercer, J.~Wakefield, A.~Pantazis, A.~Lutambi, H.~Masanja, and S.~J. Clark.
\newblock {Small area estimation of child mortality in the absence of vital
  registration}.
\newblock \emph{The Annals of Applied Statistics}, 9\penalty0 (4):\penalty0
  1889--1905, 2015.

\bibitem[Osmond and Gardner(1989)]{osmond1989age}
C.~Osmond and M.~Gardner.
\newblock {Age, period, and cohort models. Non-overlapping cohorts don't
  resolve the identification problem.}
\newblock \emph{American Journal of Epidemiology}, 129\penalty0 (1):\penalty0
  31--35, 1989.

\bibitem[Paige et~al.(2022)Paige, Fuglstad, Riebler, and
  Wakefield]{paige2022design}
J.~Paige, G.-A. Fuglstad, A.~Riebler, and J.~Wakefield.
\newblock {Design-and model-based approaches to small-area estimation in a
  low-and middle-income country context: comparisons and recommendations}.
\newblock \emph{{Journal of Survey Statistics and Methodology}}, 10\penalty0
  (1):\penalty0 50--80, 2022.

\bibitem[Papoila et~al.(2014)Papoila, Riebler, Amaral-Turkman,
  S{\~a}o-Jo{\~a}o, Ribeiro, Geraldes, and Miranda]{papoila2014stomach}
A.~L. Papoila, A.~Riebler, A.~Amaral-Turkman, R.~S{\~a}o-Jo{\~a}o, C.~Ribeiro,
  C.~Geraldes, and A.~Miranda.
\newblock {Stomach cancer incidence in Southern Portugal 1998--2006: A
  spatio-temporal analysis}.
\newblock \emph{Biometrical Journal}, 56\penalty0 (3):\penalty0 403--415, 2014.

\bibitem[Riebler and Held(2010)]{riebler2010analysis}
A.~Riebler and L.~Held.
\newblock {The analysis of heterogeneous time trends in multivariate
  age--period--cohort models}.
\newblock \emph{Biostatistics}, 11\penalty0 (1):\penalty0 57--69, 2010.

\bibitem[Riebler et~al.(2012)Riebler, Held, Rue, and Bopp]{riebler2012gender}
A.~Riebler, L.~Held, H.~Rue, and M.~Bopp.
\newblock {Gender-specific differences and the impact of family integration on
  time trends in age-stratified Swiss suicide rates}.
\newblock \emph{{Journal of the Royal Statistical Society: Series A (Statistics
  in Society)}}, 175\penalty0 (2):\penalty0 473--490, 2012.

\bibitem[Riebler et~al.(2016)Riebler, S{\o}rbye, Simpson, and
  Rue]{riebler2016intuitive}
A.~Riebler, S.~H. S{\o}rbye, D.~Simpson, and H.~Rue.
\newblock {An intuitive Bayesian spatial model for disease mapping that
  accounts for scaling}.
\newblock \emph{Statistical Methods in Medical Research}, 25\penalty0
  (4):\penalty0 1145--1165, 2016.

\bibitem[Rue and Held(2005)]{rue2005gaussian}
H.~Rue and L.~Held.
\newblock \emph{{Gaussian Markov Random Fields: Theory and Applications}}.
\newblock Chapman and Hall/CRC, 2005.

\bibitem[Rue et~al.(2009)Rue, Martino, and Chopin]{rue2009approximate}
H.~Rue, S.~Martino, and N.~Chopin.
\newblock {Approximate Bayesian inference for latent Gaussian models by using
  integrated nested Laplace approximations}.
\newblock \emph{{Journal of the Royal Statistical Society: Series B
  (Statistical Methodology)}}, 71\penalty0 (2):\penalty0 319--392, 2009.

\bibitem[Simpson et~al.(2017)Simpson, Rue, Riebler, Martins, and
  S{\o}rbye]{simpson2017penalising}
D.~Simpson, H.~Rue, A.~Riebler, T.~G. Martins, and S.~H. S{\o}rbye.
\newblock {Penalising model component complexity: A principled, practical
  approach to constructing priors}.
\newblock \emph{{Statistical Science}}, 32\penalty0 (1):\penalty0 1--28, 2017.

\bibitem[Smith(2018)]{smith2018stratified}
T.~Smith.
\newblock {A stratified age-period-cohort model for spatial heterogeneity in
  all-cause mortality}.
\newblock \emph{arXiv preprint: 1806.02748}, 2018.

\bibitem[Smith and Wakefield(2016)]{smith2016review}
T.~R. Smith and J.~Wakefield.
\newblock {A review and comparison of age--period--cohort models for cancer
  incidence}.
\newblock \emph{{Statistical Science}}, 31\penalty0 (4):\penalty0 591--610,
  2016.

\bibitem[{UN IGME}(2021)]{igme2021}
{UN IGME}.
\newblock \emph{{Levels \& Trends in Child Mortality: Report 2021}}.
\newblock {United Nations Inter-Agency Group for Child Mortality Estimation},
  \url{https://childmortality.org/wp-content/uploads/2021/12/UNICEF-2021-Child-Mortality-Report.pdf},
  2021.

\bibitem[{United Nations}(2019)]{sdgsWeb}
{United Nations}.
\newblock \emph{Sustainable Development Goals}.
\newblock \url{http://sustainabledevelopment.un.org/owg.html}, 2019.

\bibitem[USAID(2019)]{dhs}
USAID.
\newblock \emph{Demographic and Health Surveys}.
\newblock {United States Agency for International Development},
  \url{http://www.dhsprogram.com}, 2019.

\bibitem[Wakefield et~al.(2019)Wakefield, Fuglstad, Riebler, Godwin, Wilson,
  and Clark]{wakefield2019estimating}
J.~Wakefield, G.-A. Fuglstad, A.~Riebler, J.~Godwin, K.~Wilson, and S.~J.
  Clark.
\newblock {Estimating under-five mortality in space and time in a developing
  world context}.
\newblock \emph{Statistical Methods in Medical Research}, 28\penalty0
  (9):\penalty0 2614--2634, 2019.

\bibitem[Wakefield et~al.(2020)Wakefield, Okonek, and
  Pedersen]{wakefield2020small}
J.~Wakefield, T.~Okonek, and J.~Pedersen.
\newblock {Small area estimation for disease prevalence mapping}.
\newblock \emph{International Statistical Review}, 88\penalty0 (2):\penalty0
  398--418, 2020.

\bibitem[Wu and Wakefield(2022)]{wu2022modeling}
Y.~Wu and J.~Wakefield.
\newblock {Modeling Urban/Rural Fractions in Low- and Middle-Income Countries}.
\newblock \emph{{arXiv preprint: 2209.10619}}, 2022.

\end{thebibliography}

\clearpage

\end{document}